\documentclass[%
 reprint,
superscriptaddress,
 amsmath,amssymb,
 aps,%prl,
pre,
]{revtex4-2}

\usepackage{epsfig,dsfont,amssymb,amsmath,amsthm,amsfonts,amsbsy,mathrsfs}
\usepackage{graphicx}
\usepackage{color}
\usepackage{bm}% bold math
\usepackage{multirow}
\usepackage{natbib}

\newcommand{\kk}{{\mathbf k}}

\newcommand{\BEQ}{\begin{equation}}
\newcommand{\EEQ}{\end{equation}}
\newcommand{\BEA}{\begin{eqnarray}}
\newcommand{\EEA}{\end{eqnarray}}

\newcommand{\xx}{\boldsymbol{x}}
\newcommand{\yy}{\boldsymbol{y}}

\usepackage{tikz}
\usetikzlibrary{arrows,shapes,chains}

\begin{document}
\preprint{APS/123-QED}

\title{Noise-Induced Phase Separation and Time Reversal Symmetry Breaking in Active Field Theories driven by persistent noise}

%%%%%%%%%%%%%%%%%%%%%%%%%%%%%%%%%%%%%%%%%%%%%%%%%%%%%%%%%%%%%%%%%%%%%%%%%%%%%
%%

\author{Matteo Paoluzzi}
\email{matteo.paoluzzi@cnr.it}
\affiliation{Istituto per le Applicazioni del Calcolo del Consiglio Nazionale delle Ricerche, Via Pietro Castellino 111, 80131 Napoli, Italy.
}
\author{Demian Levis}
\affiliation{Departament de Física de la Mat\`eria Condensada, Universitat de Barcelona, C. Martí Franqu\`es 1, 08028 Barcelona, Spain.}
\affiliation{UBICS University of Barcelona Institute of Complex Systems, Mart\'i i Franqu\`es 1, E08028 Barcelona, Spain.}

\author{Andrea Crisanti}
\affiliation{Dipartimento di Fisica, Sapienza Universit\`a di Roma
Piazzale  A.  Moro  2,  I-00185  Rome,  Italy.}

\author{Ignacio Pagonabarraga}
\affiliation{Departament de Física de la Mat\`eria Condensada, Universitat de Barcelona, C. Martí Franqu\`es 1, 08028 Barcelona, Spain.}
\affiliation{UBICS University of Barcelona Institute of Complex Systems, Mart\'i i Franqu\`es 1, E08028 Barcelona, Spain.}

\date{\today}% It is always \today, today,
             %  but any date may be explicitly specified

\begin{abstract}
Within the Landau-Ginzburg picture of phase transitions,
scalar field theories develop 
phase separation because of a spontaneous symmetry-breaking mechanism. This picture works in thermodynamics but also in the dynamics of phase separation. Here we show that scalar non-equilibrium field theories undergo phase separation just because of non-equilibrium fluctuations driven by a persistent noise. The mechanism is similar to what happens in Motility-Induced Phase Separation where persistent motion introduces an effective attractive force.
We observe that Noise-Induced Phase Separation occurs in a region of the phase diagram where disordered field configurations 
would otherwise be stable at equilibrium. 
Measuring the local entropy production rate 
to quantify the time-reversal symmetry breaking, we find 
that such breaking is concentrated on the boundary between the two phases.
\end{abstract}
\maketitle

%\section{Introduction}
\paragraph*{Introduction.}
Dynamical field theories provide a powerful framework for investigating
collective properties in a variety of systems, ranging from critical phenomena and non-equilibrium phase transitions \cite{tauber2014critical}, to the growth of interfaces \cite{PhysRevLett.56.889}.
Continuum descriptions are also suitable 
for modeling different materials \cite{kendon2001inertial}, cell colonies \cite{cates2010arrested}, single-cell motion \cite{tjhung2012spontaneous} or  
phase transitions in cell aggregates \cite{ziebert2012model,PhysRevLett.125.038003,PhysRevE.104.054410}.
Focusing our attention on
scalar field theories, as in the case of the gas-liquid phase transition, upon introducing non-equilibrium deterministic forces, the so-called Model A and Model B
can be extended to capture the large-scale 
phenomenology of active systems, e.g., collections of self-propelled agents \cite{Marchetti13,wittkowski2014scalar,nardini2017entropy,PhysRevLett.124.240604,cates2012diffusive}. 
%Moreover, the statistical properties of the noise, 
However, also the noise fields, representing 
the effect of the fast degrees of freedom on the slow ones, 
are another 
source of non-equilibriumness in field theories \cite{sancho1998non}. 
In particular, there are no reasons a priori to consider 
delta-correlated stochastic forces,
while a natural choice might be rather to consider 
an exponential decay for the two-point correlation function 
of the noise \cite{maggi2022critical,PhysRevE.105.044139}. 
\begin{figure*}[!t]
\centering\includegraphics[width=.8\textwidth]{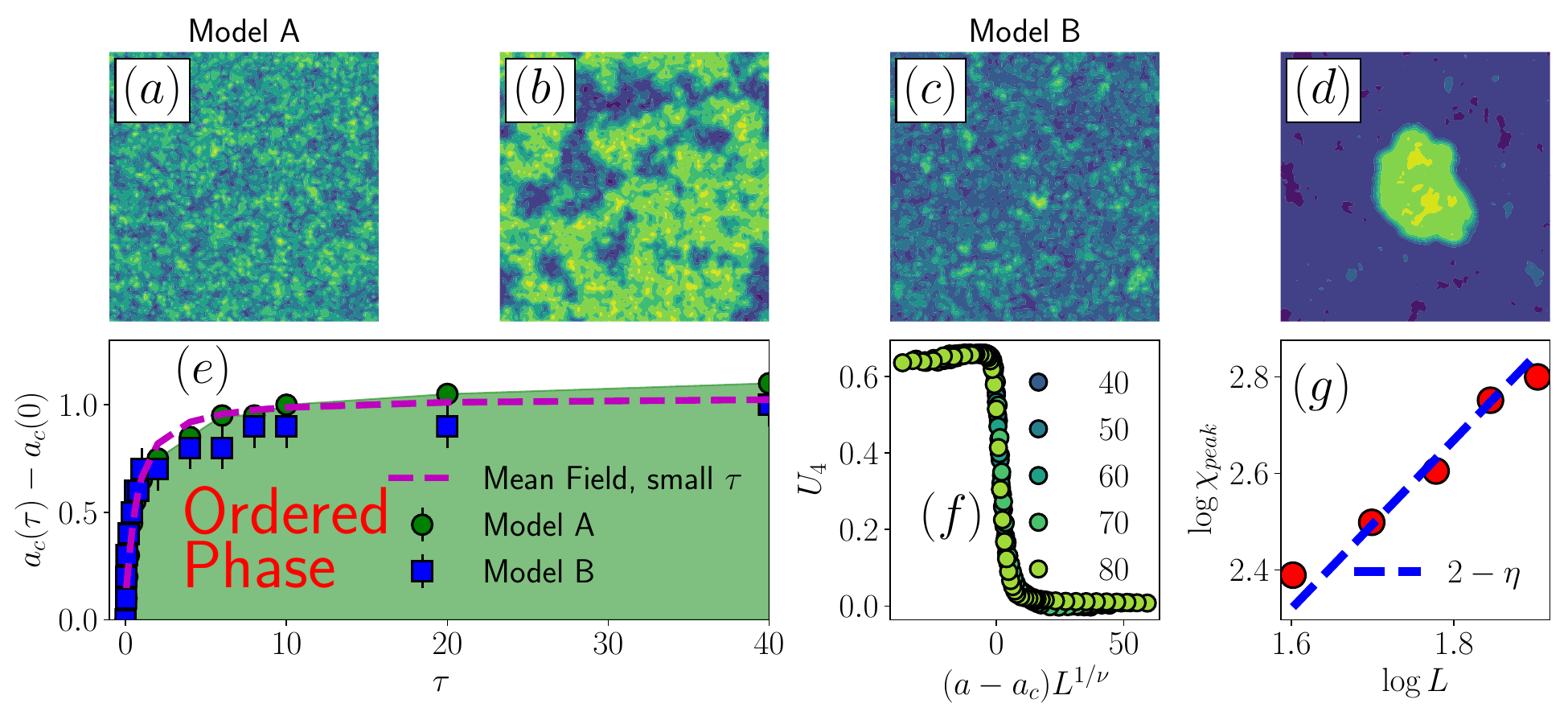}
\caption{\textbf{Noise-Induced Phase Separation (NIPS).}
(a)-(d) Stationary configurations of Model A ((a)-(b)) and Model B ((c)-(d)) in a region of the phase diagram where the equilibrium dynamics leads to a disordered phase ($a(\tau) - a(0) > 0$).  
Upon increasing $\tau$ (Model A: $\tau=0.02,2$, Model B: $\tau=0.05,10$, from left to right), 
Model A orders and Model B develops phase separation.
The lattice size is $L=100$ ($D=1$). (e) Phase diagram for Model A and B. The green region indicates where the probability distribution function $\mathcal{P}[\varphi]$ is double-peaked. The dashed magenta line is the one parameter fit to Eq. (\ref{eq:MF}).  (f) Rescaled Binder parameter for $\tau=1$ with $\nu=1$ and different system sizes from $L=40$ to $L=80$ (see legend). (g) Scaling of the susceptibility for $\tau=1$ is consistent with $\eta=1/4$. 
}
\label{fig:fig1}      
\end{figure*}

In this work, we study
non-conserved (Model A) and conserved (Model B) 
scalar field theories in $2d$, driven out-of-equilibrium 
by time-correlated noise.  
We document that the non-equilibrium 
noise is the driver of phase separation in a region of the phase diagram where the corresponding equilibrium system does not display any ordered phase. 
Because the effect of the persistent noise is to destabilize homogeneous configurations as in the case of self-propulsion in active systems (the so-called Motility-Induced Phase Separation (MIPS) \cite{Tailleur08}), we call this phenomenon Noise-Induced Phase Separation (NIPS). However, distinct from the coarse-grained theories of MIPS, NIPS does not require non-equilibrium terms for breaking time-reversal symmetry (TRS)  or to produce micro-phase separation \cite{nardini2017entropy}. 
As in the case of MIPS, but without any local non-equilibrium terms in the deterministic force, TRS breaking (TRSB), measured using the entropy production rate, is concentrated at the interface between the two phases.
\paragraph*{ Correlated Noise \& Active Field Theories.}
Models with exponentially correlated noise have been largely employed during the last years in Active Matter \cite{szamel2014self,maggi2015multidimensional,Farange15,Fodor16}. 
Experiments show that active baths are a source of exponentially correlated noise \cite{Maggi14,maggi2017memory}. 
This picture holds even in dense living materials \cite{henkes2020dense}.
Numerical simulations show that the leading order dynamical field theory describing the MIPS critical point is driven by a correlated noise \cite{maggi2022critical}. To provide a concrete example of 
how the echo of the activity takes the form of a persistent noise on the mesoscopic scale, we consider the practical situation where active particles are in contact with a thermal bath. Even though the effect of thermal diffusion is negligible (for instance, in the case of E. coli bacterial baths thermal diffusion is almost two orders of magnitude smaller than diffusivity due to activity \cite{Maggi14}), once we take into consideration thermal fluctuations, we can coarse-grain the dynamics with respect these fast degrees of freedom, keeping the active force fixed. This is still well-justified by experiments: again, in the case of E. coli, the relaxation time of thermal degrees of freedom at room temperature is of the order of $\tau_{T} \sim 10^{-7}s$ while the persistence time %, and thus the crossover between active ballistic regime and active diffusive regime, 
  is of the order of $\tau_A \sim 1 s$. This number decreases with bacterial density, but it does not change the orders of magnitude so that we are always in the situation $\tau_A \gg \tau_T$. In SI we report the explicit coarse-graining of a model of Active Ornstein-Uhlenbeck particles with active and thermal noise showing that the adiabatic average of the thermal fluctuations brings to field theories with correlated noise.
In the following, we will explore the implication of this effect on $2d$ scalar field theories.

\paragraph*{Model.}
We consider the relaxation 
dynamics of a scalar field $\varphi\equiv\varphi(\xx,t)$.
$\varphi$ represents the slow variable we are interested in,
as in the case of density fluctuations in the gas-liquid phase transition.
The dynamics of $\varphi$ results from the competition between a deterministic force $F$ and a fluctuating one $\psi\equiv\psi(\xx,t)$. The latter represents 
the effect of the fast degrees of freedom on $\varphi$. Instead of considering the 
stochastic force as a zero mean and delta correlated noise, we assume $\psi$ to be an
annealed variable characterized by a time scale $\tau$ which is another control parameter of the model (in the limit $\tau \to 0$, we recover a delta-correlated noise). 
The dynamics reads
\begin{align} \label{eq:dynamics}
    \dot{\varphi} = F[\varphi] + \psi
\end{align}
with $\langle \psi \rangle \!=\!0$ and $\langle \psi(\xx,t) \psi(\yy,s) \rangle \!=\! 2 D L( \xx \!-\! \yy , t  \!-\! s)$,
where $D$ sets the strength of the noise and the operator $L$ keeps into account both, a suitable differential operator for describing conserved or non-conserved dynamics, and the time-correlation function of the noise. The expressions for $F$ and $L$ are reported in (\ref{tab:modA/B}). 
We restrict our study to the case where $F$ can be written as the functional derivative of the standard Landau-Ginzburg energy functional 
\begin{align} \label{eq:LG}
    H_{LG}[\varphi] = \int d\xx \, \left[ \frac{\mu}{2} (\nabla \varphi)^2 +\frac{a}{2} \varphi^2 + \frac{u}{4} \varphi^4 \right] \; ,
\end{align}
so that non-equilibrium is caused solely by the presence of the time-correlated noise $\psi$. The parameter $a$ sets the distance from the equilibrium critical point $a\!=\!0$ \footnote{Below the upper critical dimension, this critical value will be renormalized to lower values because fluctuations are not negligible. 
}. 
\begin{figure}[!t]
\centering\includegraphics[width=.4\textwidth]{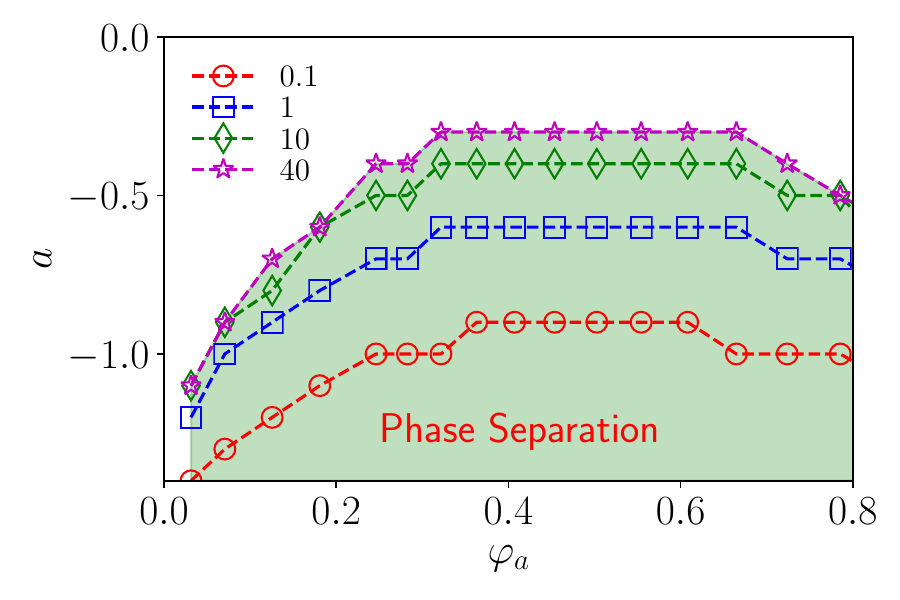} %\\
%\centering\includegraphics[width=.4\textwidth]{Figure2new_upper.pdf} 
\caption{ \textbf{Non-equilibrium dynamics enhances phase separation.} Phase diagrams of Model B for different values of $\tau$ (see legend). The green area indicates the phase separation region (corresponding to $\tau\!=\!40$) that increases monotonically for increasing values of $\tau$ (see \cite{SM}).}
%(b)-(e) Phase-separated configurations ($\tau=5,a=-2$) for different values of $\varphi_a=0.01,0.09,0.25,0.36,0.64$, from left to right.
%} 
\label{fig:fig2}      
\end{figure}

\begin{table}
    \centering
    \begin{tabular}{c|c|c|}
                & $F[\varphi]$ & $L(\xx,t)$ \\ \hline
       Model A  & $-\frac{\delta H_{LG}}{\delta \varphi}$ & $\delta( \xx ) K(t)$ \\
       Model B  & $\nabla^2 \frac{\delta H_{LG}}{\delta \varphi}$ & $-\nabla^2 \delta(\xx) K(t)$ \\
    \end{tabular}
    \caption{ Definitions of $F$ and $L$ for Model A and B.}
    \label{tab:modA/B}
\end{table}

\paragraph*{Phase diagram \& NIPS.} 
We start with
the mean-field (MF) picture
within an effective equilibrium approach. For convenience, we consider 
the Markovian approximation named Unified Colored Noise (UCN) \cite{Hanggi95}, it has been shown that the correlated noise shifts the critical point of the Landau model at higher temperatures, i.e, the critical value $a_c(\tau)$ is in general $a_c(\tau) > a$ \cite{Paoluzzi16}.  
Although UCN dynamics does not reproduce the real trajectories \cite{PhysRevE.107.034110}, it provides useful insight into the stationary properties of the system, especially when the potential generating deterministic forces are characterized by positive curvatures, as in the case of a $\varphi^4$ theory approaching the critical point from above. For moving at higher order in $\tau$, we should employ other perturbation schemes whose leading order in $\tau$ matches our computation \cite{Fodor16,PhysRevE.103.032607}.
In the small-$\tau$ regime, one has 
$H_{eff}\!\simeq\!H_{LG} \!+\! \frac{\tau}{2}(H_{LG}^\prime)^2 \!-\! \tau H^{\prime\prime}_{LG}$, where the prime indicates the derivative with respect to $\varphi$. A phase transition to $\varphi \!\neq\! 0$ takes place if the configuration $\varphi\!=\!0$ is not stable anymore. We can check the stability of $\varphi\!=\!0$ looking at the second derivative of $H_{eff}$, given by  $H^{\prime\prime}_{eff} \!=\! H_{LG}^\prime + \tau \left[ (H_{LG}^{\prime \prime})^2 + H_{LG}^\prime H_{LG}^{\prime \prime\prime }\right] - \tau H_{LG}^{\prime \prime\prime \prime} $. Since $H^{\prime\prime}_{eff}[0]\!=\!-3u\tau$, negative for any $\tau\!>\!0$,  the system undergoes a phase transition to $\varphi \!\neq\! 0$. Since the parameter driving the phase transition is the non-equilibrium noise, we refer to this mechanism as NIPS. 

The stationary homogeneous configurations of $\varphi$ are thus regulated by \cite{SM}
\begin{align} 
    H_{eff} &=\frac{\tilde{a}}{2} \varphi^2 + \frac{\tilde{u}}{4} \varphi^4 + O(\varphi^6) \\ \nonumber 
    \tilde{a} &\equiv a (1 + \tau a) - 3 u \tau\,, \  
    \tilde{u}\equiv u (1 + 4 \tau) \; . 
\end{align}
We obtain that the combination of non-equilibrium noise, parametrized by $\tau$, and non-linear interactions, whose intensity is tuned by $u$, renormalizes the coupling constants $a$ and $u$,
i.e., $\tilde{u} \!\geq\! u$ so that the non-linear interaction becomes more important, and $\tilde{a} = \tilde{a}(\tau, u)$, so that it can change its sign. 
Without non-linear interactions ($u\!=\!0$), 
the location of the transition remains untouched at $a\!=\!0$. 
In the language of quantum field theory, $a$ represents the mass of the scalar field, and thus the mechanism introduced here predicts that, just because of the interaction with the external annealed field $\psi$ of mass $m_\psi = \tau^{-1}$, the field $\varphi$ can acquire mass whose value is proportional to $m_\psi$.
For $u\!>\!0$,
the shift of the critical point (or the mass of the scalar field $\varphi$) is given by 
$\tilde{a}\!=\!0\!\equiv\! a_c(\tau)$, that is
\begin{align} \label{eq:MF}
    a_c(\tau) = \frac{1}{2 \tau} \left( \sqrt{1 + 12 u \tau^2} -1 \right) \; .
\end{align}
Away from the MF regime (and out from the effective equilibrium), 
it is not clear how 
fluctuations and non-equilibrium dynamics 
change the MF. 

We now move to $2d$ using numerical simulations. As in MF, we observe that the correlated noise drives a phase transition. This is qualitatively shown in Fig. (\ref{fig:fig1}) where we report representative configurations of Model A ((a)-(b)) and B ((c)-(d)) 
away from phase transitions at $\tau=0$.
As one can see, for increasing values of $\tau$,
disordered configurations become unstable so that a phase transition takes place. 
To make quantitative progress, we compute the phase diagram of the model in both cases, Model A and Model B. The result is shown in Fig. (\ref{fig:fig1}e).
We observe that by tuning $\tau$ the system undergoes a non-equilibrium phase transition in a wide region of the phase diagram where the corresponding equilibrium system is homogeneous. 
We compare the numerical results with the theoretical prediction (\ref{eq:MF}) using $u$ as a fitting parameter 
(this is because $u$ in (\ref{eq:MF}) is the MF value and not the one renormalized by fluctuations). 
As one can see, the theory reproduces remarkably well the order-disorder transition in a wide range of $\tau$ (with $u_{fit}\simeq 0.1$). 
This is counterintuitive since eq. (\ref{eq:MF}) has been obtained within the small-$\tau$ limit. We can rationalize this by noticing that  
$\varphi$ in the Landau theory of continuous transition is 
arbitrary small
at the transition so that the correction 
to the mass $\tau H^{\prime \prime}_{LG} \simeq \tau u \varphi^2$ around the critical point is small in a wide range of $\tau$.

To provide an estimate of the critical exponent $\nu$ of Model A, 
we now compute the Binder parameter $U_4 
\!\equiv\! 1 - \langle \varphi^4 \rangle / 3 \langle \varphi^2 \rangle^2$ for system sizes $L\!=\!40,50,60,70,80$ and $\tau\!=\!1$ 
(see \cite{SM} for details). 
As shown in Fig. (\ref{fig:fig1}f), we observe a good scaling collapse within the Ising universality class, i.e., $\nu\!=\!1$. Measuring the exponent $\eta$ through the scaling of the peak of the susceptibility $\chi$, we observe again a value consistent with Ising universality class (Fig. (\ref{fig:fig1}g)).  

\begin{figure}[!t]
\centering\includegraphics[width=.4\textwidth]{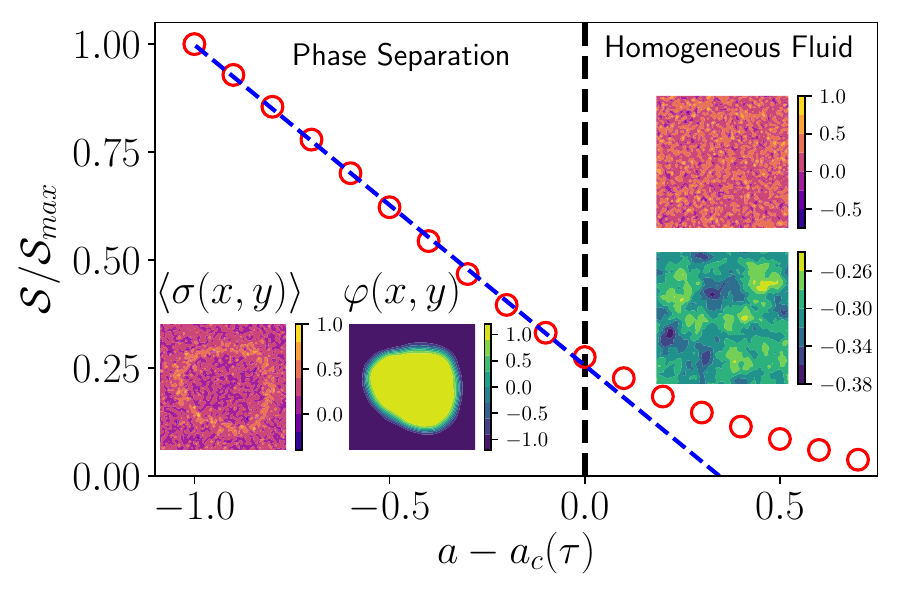}
\caption{ \textbf{Entropy production rate and phase separation.}  Total entropy production rate $\mathcal{S}$ as a function of the distance from the transition point, i.e., $a_c(\tau) - a$, in the case of Model B ($\tau\!=\!1$). $\mathcal{S}$ linearly increases in the phase-separated regime (see the blue dashed line as a guide to the eye).
Insets: Representative stationary configuration of the conserved field $\varphi$ and the corresponding local entropy production rate $\sigma$.}
\label{fig:fig3}      
\end{figure}

Next, we measure how non-equilibrium fluctuations modify the phase separation region in Model B.
We thus performed numerical simulations for $\tau=0.1,1,10,40$ and several values of the initial density $\varphi_a \equiv \int d\boldsymbol{x} \, \varphi(\boldsymbol{x},0)$. Fig. (\ref{fig:fig2}) reports 
the phase-separation region as $\tau$ increases.
We observe that not only $\tau$ is the trigger for the phase separation, but it also quantitatively impacts the size of the phase-separated region making it larger and larger for increasing values of $\tau$ (the scaling of the size of the phase separation region with $\tau$ is shown in \cite{SM}).

\begin{figure}[!t]
\centering\includegraphics[width=.5\textwidth]{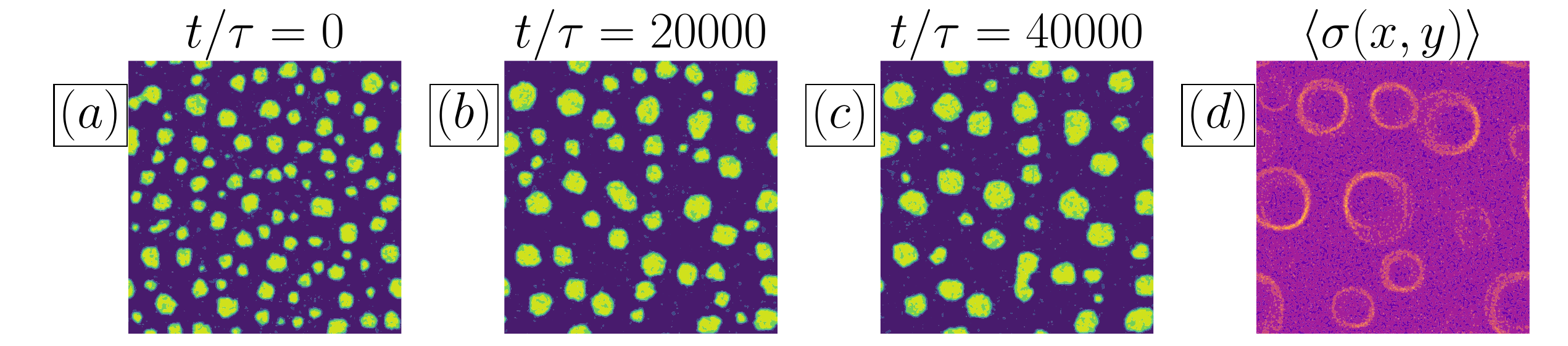} \\
\centering\includegraphics[width=.5\textwidth]{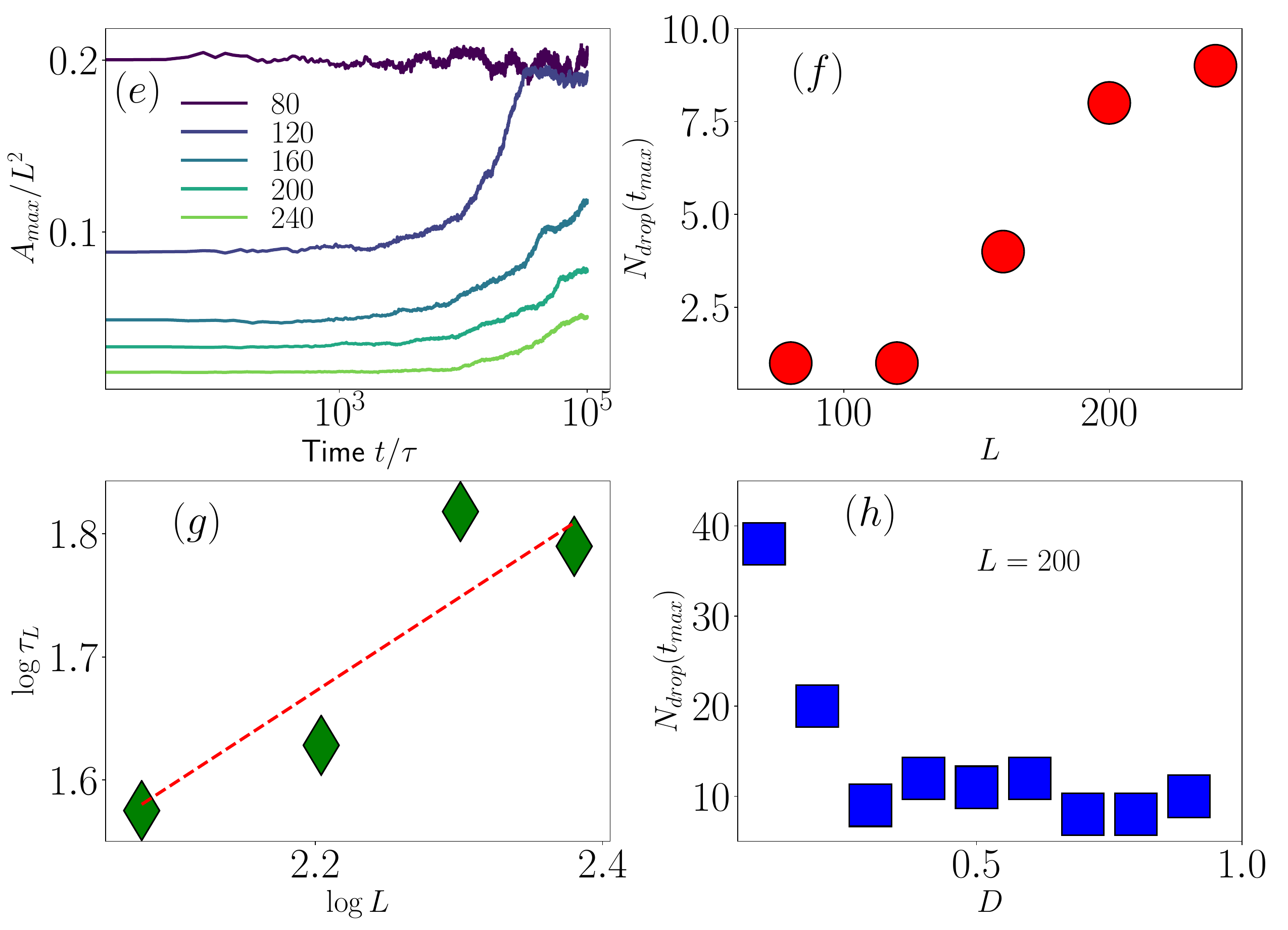}
\caption{ \textbf{Microphase separation. (a)--(c)} Phase-separated configurations ($\tau=0.5,a=-3$) for the larger system size $L=240$ at different times. (d) Entropy Production Rate Field. (e) Drop area as a function of time for different system sizes (see legend). (f) Number of drops in the final configuration as a function of the system sizes. (g) Relaxation time $\tau_L$ of microphase as a function of the system size $L$ (the dashed line power law scaling is a guide to the eye). 
(h) Number of drops as a function of the noise strength $D$.}
\label{fig:fig4}      
\end{figure}

\paragraph*{TRSB. }
As NIPS is driven by non-equilibrium fluctuations, a natural question 
is whether the noise is just a trigger 
for an equilibrium-like transition. 
To answer this question we look at the total entropy production rate $\mathcal{S}$. 
%which is the natural observable to quantify TRSB. \
$\mathcal{S}$ has been the subject of intense studies during the last decades for rationalizing the thermodynamics of Active Matter \cite{bebon2024thermodynamics,fodor2022irreversibility,PhysRevX.11.021057,borthne2020time}. For our purpose, $\mathcal{S}$ is a proxy for measuring the breakdown of Time Reversal Symmetry
% whether or not the field theory breaks TRS 
\cite{PhysRevX.11.021057,borthne2020time}.
$\mathcal{S}$ is defined as the long-time behavior of Kullback-Leibler divergence  \cite{lebowitz1999gallavotti}
\begin{align}
    \mathcal{S} \equiv \lim_{T \to \infty} \frac{1}{T} \left\langle \log \frac{P[\varphi]}{P_{R}[\varphi]} \right\rangle \; .
\end{align} with $P[\varphi]$ indicating the probability of the path $\varphi(\xx,t)$ with $t \in[t_0,T]$ and $P_R[\varphi]$ the probability of the time-reversed path $\varphi_R$ obtained through the transformation $t \to T - t$. 
In the case of a field theory, $\mathcal{S}$ can be written in terms of a 
spatial resolved entropy production rate $\sigma(\boldsymbol{x})$ so that $\mathcal{S} \!=\! \int d\boldsymbol{x} \, \sigma(\boldsymbol{x})$ where $\sigma(\boldsymbol{x})$ is a model-dependent composite operator of the field $\varphi$ \cite{nardini2017entropy,PhysRevE.105.044139,pruessner2022field}. 
In the case of Model B, the computation of $\mathcal{S}$ brings to \cite{SM}
\begin{align} \label{eq:epr_main}
\sigma(\boldsymbol{x}) = \frac{\tau^2}{2 D} \left\langle \dot{\varphi}^3 \frac{\delta^3 H_{LG}}{\delta \varphi^3}\right\rangle 
\end{align}
where the angular brackets indicate the averaging over time performed on a stationary configuration (the presence of the cutoff $\Lambda=2 \pi / \ell$, with $\ell$ the lattice spacing, avoids any ultraviolet divergence). We highlight that the expression we arrive at is the same obtained earlier for Model A \cite{PhysRevE.105.044139} and in many-body \cite{Fodor16,PhysRevE.103.032607}. 

We employ (\ref{eq:epr_main}) for computing $\mathcal{S}$ in simulations. Fig. (\ref{fig:fig3}a) reports $\mathcal{S}$ ($\tau\!=\!1$).
As an initial condition, we consider a drop of radius $R\!=\!20$ ($L\!=\!60$). We observe that $\mathcal{S}$ increases approaching
the transition. As in other non-equilibrium field theories \cite{alston2023irreversibility,suchanek2023irreversible}, $\mathcal{S}$ undergoes a crossover at the transition where it starts to grow linearly for decreasing value of $a$. 
Because $\mathcal{S}$ is non-zero, despite the phase separation being equilibrium-like, 
TRS remains broken for maintaining
 the two phases separated.
 
Another natural question is whether TRSB is accompanied by some non-equilibrium pattern formation.  
We thus look at the map  
$\sigma(\boldsymbol{x})$. In Fig. (\ref{fig:fig3}) we display the field configurations and the corresponding $\sigma(\boldsymbol{x})$ (see the insets). In the symmetric phase, $\varphi$ is disordered in space and the same happens for $\sigma$. Once the system phase separates, 
the corresponding map of $\sigma$ develops a pattern at the boundary between the two phases, indicating that most TRSB is concentrated in that region. This is precisely the kind of pattern 
observed in experiments and simulations of Active Systems in a model-independent fashion \cite{PhysRevLett.129.220601}.
We can rationalize that from (\ref{eq:epr_main}) noticing that terms proportional to $\varphi^m \nabla^2 \varphi$, with $m > 1$, that give a contribution to $\sigma$ on the boundary between the two phases, arise once we replace $\dot{\varphi}$ by eq. (\ref{eq:dynamics}), so that $\sigma \sim \langle (\mu \nabla^2 \varphi - a\varphi + u \varphi^3 + \psi  )^3 \varphi \rangle$. These terms are irrelevant in the Renormalization Group sense \cite{PhysRevE.105.044139}, however, similarly to the case of Active Model B (but distinct since we do not have any non-equilibrium gradient term in $F$), they contribute to a space-dependent TRSB. In the case of Model A, the order parameter is not conserved and thus interfaces are suppressed in favor of homogeneous phases that spontaneously break the $\varphi \!\to\! -\varphi$ symmetry. 
$\sigma$ might develop anomalous fluctuations around the upper critical dimensions $d_c\!=\!4$ \cite{PhysRevE.105.044139,PhysRevLett.124.240604}, which is far away from our numerical study.
\paragraph*{Microphase separation \& Coarsening.} The non-equilibrium noise 
impacts the morphology of the phase separation: 
 we observe that, deep in the phase-separated region, 
 Model B develops microphases \cite{stenhammar2014phase, caporusso2020motility}, as shown in Fig. (\ref{fig:fig4})a-c ($L\!=\!240$, $ a\!=\!-3$, and $\tau\!=\!0.5$) where we report 
 typical  
 configurations at different times for $L=240$ \cite{SM}. 
 In Fig. (\ref{fig:fig4})d we show $\sigma$ that signal TRSB along the edge of each drop. Performing longer simulations ($N_t = 1.2 \times 10^7$ steps) for several system sizes ranging from $L=80$ up to $L=240$ we document a crossover from microphase separation to coarsening. Fig. (\ref{fig:fig4})e shows how the area covered by the larger drop $A_{max}$ evolves for different system sizes (we wait $2 \times 10^6$ steps before collecting data, $\tau=0.5$ and $a=-3$). 
 The time spent in configurations with small drops, i.e., the larger drop does not exceed the $10 \%$ of the box area, increases with $L$. In other words, after developing microphases, the system undergoes a coarsening dynamics towards a single large droplet after decreasing the system sizes. In the thermodynamic limit, the system never reaches the coarsening regime but remains microphase separated. This is documented also in (f) where we report the number of drops in the final configuration. We thus measure the characteristic lifetime $\tau_L$ of the microphase by fitting to an exponential decay the number of drops $N_{drops}(t)$ as a function of time. The result of this analysis is reported in (g) for different system sizes $L$. We obtain that $\tau_L$ increases linearly with $L$. Finally, we explore the effect of the strength of the noise $D$ (see Fig. (\ref{fig:fig4})h)
 study how the strength of the noise $D$ impacts microphases obtaining that the number of drops increases significantly as $D$ decreases, as in the case of Active Model B+ \cite{PhysRevX.8.031080}.

\paragraph*{Discussion.} We have shown that scalar field theories in MF and $2d$ 
undergo a phase transition driven by non-equilibrium noise and controlled by its persistence time $\tau$.
The phase separation occurs in a region of the phase diagram where the corresponding equilibrium model is homogeneous. 
We showed that the emerging phenomenology can be rationalized within a simple MF theory within the small-$\tau$ limit. 
The theory highlights how the combination of non-linearities (parametrized by $u$) and non-equilibrium noise (parametrized by $\tau$) can trigger the transition.
We computed numerically the critical exponents that are compatible with the Ising universality class \footnote{The accurate numerical computation of the critical 
exponents requires separate work.}. The numerical computation of $\mathcal{S}$ indicates that NIPS breaks TRS making the phase separation distinct from that of the equilibrium model. Again, TRSB is due to the combination of non-linear interactions and non-equilibrium dynamics: both ingredients are fundamental and complementary.
In other words, even though the critical point 
belongs to the Ising universality class, 
phase separation is maintained because of the continuous energy injection on the mesoscopic scale due to the noise.
Because of that, $\mathcal{S}\neq 0$ 
 for any arbitrary small value of $\tau$. 
Finally, we observed that most of the TRSB is concentrated at the boundaries between the two coexisting phases. It is worth noting that the picture we obtain is quite similar to that of Active Model B. 
On the other hand, in our model, TRSB is caused by the presence of non-equilibrium fluctuations that couple with each other because of non-linear interactions. We also documented other features of scalar Active Matter that usually require extra non-equilibrium terms in the framework of Active Model B, such as microphase separation \cite{PhysRevX.8.031080,caballero2018bulk,PhysRevLett.127.068001, PhysRevLett.125.168001}.

Finally, we observe that time-correlated noise naturally emerges in the continuum description of Active Matter. For instance, the set of continuum equations usually takes the form $\partial_t \rho \!=\! -\nabla  (J -  D(\rho) \nabla \rho)$  with a current $J$  decaying on a finite time-scale, i.e., $\partial_t J \!=\! -D_r J + \dots$ \cite{speck2015dynamical,Marchetti13,PRLMarchetti2012,marconi2021hydrodynamics}. 
To perform the linear stability analysis one considers $J$ 
as a fast variable so that it can be removed adiabatically
$\dot{J}\!=\!0$. If we remove this assumption, $\nabla J$ acts as an Ornstein-Uhlenbeck field on $\rho$. 
Our results suggest that, once we include non-linear interactions, the Ornstein-Uhlenbeck field destabilizes homogenous profiles, even in the small (but not vanishing) $\tau$ regime.

\paragraph*{Acknowledgments.}
M.P. acknowledges NextGeneration EU (CUP B63C22000730005), Project IR0000029 - Humanities and Cultural Heritage Italian Open Science Cloud (H2IOSC) - M4, C2, Action 3.1.1.
D.L. acknowledges  DURSI and AEI/MCIU for financial support under project 2021SGR00673 and PID2022-140407NB-C22.
I.P. acknowledges MICINN, DURSI, and SNSF for financial support under
Projects No. PGC2018-098373-B-I00, No. 2017SGR-884,
and No. 200021-175719, respectively.

\bibliography{mpbib}
\bibliographystyle{rsc}

\clearpage
\newpage
\onecolumngrid
\renewcommand{\thefigure}{S\arabic{figure}}
\renewcommand{\theequation}{S\arabic{equation}}
\setcounter{figure}{0}
\section*{Supplemental Information}

\subsection*{Shift of the critical temperature in the small-$\tau$ limit}

In order to illustrate how to compute the shift of the critical temperature in the small $\tau$ limit (further details are provided in Ref. \cite{Paoluzzi16}), we start from a zero-dimensional Landau model driven out-of-equilibrium by an exponentially correlated noise that can be written in terms of an Ornstein-Uhlenbeck process so that
\begin{align} \label{eq:dynMF}
    \dot{\varphi} &= -\frac{\partial H_{LG}}{\partial \varphi} + \psi \\ \nonumber
    \dot{\psi} &= -\tau^{-1} \psi + \xi \\ \nonumber
    H_{LG} &= \frac{a}{2} \varphi^2 + \frac{u}{4} \varphi^4  \; ,
\end{align}
with $\langle {\xi}(t) {\xi}(s)\rangle = 2 D \tau^{-2} \delta(t-s)$.
The steps that bring to the effective equilibrium picture (details can be found in Ref. \cite{Hanggi95}) are the following
\begin{itemize}
    \item We reduce the dynamics to a single equation for $\ddot{\varphi}$ that is
    \begin{align}
        \tau \ddot{\varphi} + \Gamma(\varphi) \dot{\varphi} &= F[\varphi] + \hat{\xi} \\ \nonumber 
        F[\varphi] &\equiv -\frac{\partial H_{LG}}{\partial \varphi} \\ \nonumber 
        \Gamma(\varphi) &\equiv 1 + \tau \frac{\partial^2 H_{LG}}{\partial \varphi^2}\; .
    \end{align}
    with $\langle \hat{\xi}(t) \hat{\xi}(s)\rangle = 2 D \delta(t-s)$.
\item In the following, we assume to approach the transition from the disordered phase so that we do not incur any problem with negative friction $\Gamma(\varphi)$.
\item We perform the large friction limit $\Gamma(\varphi) \to \infty$ so that we can neglect the inertial term 
\begin{align}
    \Gamma(\varphi) \dot{\varphi} \simeq F[\varphi] + \xi 
\end{align}
\item The corresponding Fokker-Planck solution 
for the probability $P[\varphi,t]$ 
has stationary solution 
$\lim_{t \to \infty} P[\varphi,t] = P_{st}[\varphi ]$ 
equilibrium-like (without loss of generality, we now set $D=1$)
\begin{align}
    P_{st}[\varphi ] &= \mathcal{N} \times e^{-H_{eff}[\varphi]} \\ \nonumber
    H_{eff}[\varphi] &\equiv H_{LG} + \frac{\tau}{2} (\frac{\partial H_{LG}}{\partial \varphi})^2 - \log \Gamma[\varphi] \; .
\end{align}
\item We now employ a small-$\tau$ limit
\begin{align}
    H_{eff}[\varphi] &\simeq H_{LG} + \frac{\tau}{2} (\frac{\partial H_{LG}}{\partial \varphi})^2 - \tau \frac{\partial^2 H_{LG}}{ \partial \varphi^2}\; .
\end{align}
\item Once we collect the different terms, we obtain
\begin{align}
    H_{eff} [\varphi] &= \frac{\varphi^2}{2} \left( a + \tau a^2 - 3 u \tau \right) + \frac{\varphi^4}{4} \left( u + 4 \tau u\right) + \frac{\tau u^2}{2} \varphi^6 - \tau a \; .
\end{align}
\item We neglect the constant term $-\tau a$ and higher order $\varphi$ terms (we notice that for $\tau>0$ we always have a quartic term meaning that the term $\varphi^6$ does not play any role) 
\begin{align}
    H_{eff} &=\frac{\tilde{a}}{2} \varphi^2 + \frac{\tilde{u}}{4} \varphi^4 + O(\varphi^6) \\ \nonumber 
    \tilde{a} &\equiv a (1 + \tau a) - 3 u \tau \\ \nonumber 
    \tilde{u} &\equiv u (1 + 4 \tau) \; . 
\end{align}    
\end{itemize}
We now compute the transition point of the effective model by looking at the point where theory becomes massless, i.e., by solving the equation $\tilde{a}=0$ whose solution brings to (9) of the main text.

The mean-field picture discussed here can be obtained considering a field theory driven by exponentially correlated noise once we (i) perform the small-$\tau$ limit, and (ii) we neglect fluctuations. 
Working at the first-order in $\tau$, our strategy is to work within Unified Colored Noise Approximation (UCN) and thus move to $\tau \to 0$. It is worth noting that for moving our analysis at the highest $\tau$-order, both UCN and small-$\tau$ approximations are both not suitable and perturbative schemes in terms of $\tau^{1/2}$ are   more appropriate \cite{Fodor16,PhysRevE.103.032607}. However, considering only corrections of the order of  $\tau$, all the computations schemes produce the same result. 
In particular, the mean-field description we have considered (that is based on a single degree of freedom), can be obtained by considering the general field theory
\begin{align}
    \dot{\varphi}(x,t) = F[\varphi(x,t)] + \eta(x,t)
\end{align}
and thus move to the Fourier space (see the next section for details) so that we obtain a many-body-like set of equations
\begin{align}
    \dot{\varphi}_k(t) = F_k + \eta_k(t)\, .
\end{align}
At this level, $F_k$ takes the form
\begin{align}
    F_k = -\frac{\partial H_{LG}}{\partial \varphi_k} \; .
\end{align}
In Fourier space $H_{LG}$ reads (see the next section for further details)
\begin{align}
    H_{LG} = \sum_{k} \left[k^2 + a \right] \varphi_k^2 + H_{I} \; .
\end{align}
The mean-field description can be obtained once we consider $k \to 0$ so that we get rid of any gradient terms. Once we do that, we end up with a single non-linear oscillator driven by a correlated noise that can be described using Eq. \ref{eq:dynMF}.

\subsection*{Entropy Production Rate of non-equilibrium Model B}
We start with the equation of motion of the conserved scalar field which is 
\begin{align}
    \dot{\varphi} &= F[\varphi] + \psi \\ 
    F[\varphi] &\equiv \nabla^2 \frac{\delta H_{LG}}{\delta \varphi}
\end{align}
and the two-point function of the noise reads
\begin{align}
    \langle \psi(\boldsymbol{x},t) \psi(\boldsymbol{y},s)\rangle &= - 2 D \nabla^2  \delta(\boldsymbol{x} - \boldsymbol{y}) K(t-s) \equiv 2 D L(\boldsymbol{x} - \boldsymbol{y}, t - s) \\ 
    L(\xx,t) &\equiv -\nabla_{\xx}^2 \delta(\xx) K(t) \\ 
    K(t) &\equiv \frac{1}{\tau} \, e^{-|t| / \tau} \; .
\end{align}
As shown in \cite{PhysRevE.107.034110}, the inverse operator $K(t)$ is
\begin{align}
    K^{-1}(t,s) = \left[ -\tau^2 \frac{d}{dt^2} + 1\right] \delta(t-s) \; .
\end{align}

Once we Fourier transform in space the fields
\begin{align}
    \varphi &= \frac{1}{L^{d/2}} \sum_{ |\kk| \in (0,\Lambda)} e^{i \kk \cdot \xx} \varphi_\kk(t) \\
    \psi    &= \frac{1}{L^{d/2}} \sum_{ |\kk| \in (0,\Lambda)} e^{i \kk \cdot \xx} \psi_\kk(t) 
\end{align}
where the ultraviolet cutoff $\Lambda=\frac{2 \pi}{\ell}$ regularizes any ultraviolet divergence, with $a$ the lattice spacing $a$. Once we Fourier transform the fields,
we obtain the following set of equations of motion
\begin{align}
    \dot{\varphi}_\kk &= F_{\kk} + \psi_\kk 
\end{align}
where we have defined $F_{\kk}$ as
\begin{align}
    F_{\kk} &= -k^2 \frac{\partial H_{LG}}{\partial \varphi_\kk} \\ 
    H_{LG} &= H_0 + H_I \\
    H_0 &\equiv \frac{1}{2} \sum_{|\kk| \in (0,\Lambda)} \left[ \mu k^2 + a\right] |\varphi_k|^2\\ 
    H_I &\equiv \frac{u L^d}{4 !} \sum_{ \kk_1, \kk_2, \kk_3, \kk_4} \varphi_{\kk_1} \varphi_{\kk_2} \varphi_{\kk_3} \varphi_{\kk_4} \delta_{\kk_1 +\kk_2+\kk_3+\kk_4}
\end{align}
and the two-point function of the time-correlated noise reads
\begin{align}
    \langle \psi_\kk(t) \psi_{\kk^\prime} (s)\rangle = 2D \, k^2 \delta_{\kk + \kk^\prime} K(t-s)\; .
\end{align}
After standard manipulations, we arrive at the following expression for the probability of a path
\begin{align}
    P[\varphi] &\propto \exp{ \left\{ -\mathcal{A}[\varphi] \right\} } \\ 
    \mathcal{A}[\varphi] &= \int_{t_0}^{t_F} \, dt \left( \mathcal{L}^{TR}[\varphi] + \mathcal{L}^{TRB}[\varphi]\right) + B.T. + T.R.T. \\ 
    \mathcal{L}^{TR}[\varphi] &\equiv  \sum_k^\prime \frac{1}{4 D k^2} \left[\dot{\varphi}_k + F_k \right] \\
    \mathcal{L}^{TRB}[\varphi] &\equiv -\sum_{k,l,m}^\prime \frac{\tau^2}{4 D} \dot{\varphi}_k \dot{\varphi}_l \dot{\varphi}_m \frac{\partial^3 H_I}{\partial \varphi_k \partial \varphi_l \partial \varphi_m} \; . 
\end{align}
where the prime in the sum indicates that it is limited by the ultraviolet cutoff $\Lambda$.
$B.T.$ indicates boundary terms that depend on the field configurations at $t_0$ and $t_F$, and $T.R.T.$ indicates other terms that do not break the time-reversal symmetry.  $\mathcal{L}^{TR}[\varphi]$ is the standard Onsager-Machlup Lagrangian density that, since there are no non-integrable terms in $F_k$, it is time-reversal and thus it does not contribute to the entropy production rate. $\mathcal{L}^{TRB}$ breaks the time-reversal symmetry as soon as $\tau > 0$.

The total entropy production rate $\mathcal{S}$ is defined as
\begin{align} \label{eq:epr}
\mathcal{S} &= \lim_{t_F \to \infty } t_F^{-1} \left\langle \log \frac{P[\varphi]}{P_R[\varphi]} \right\rangle \; ,
\end{align}
where the probability of the time-reversal path $P_R[\varphi]$ can be computed considering the time-reversal transformation $t^\prime = t_F - t$ so that we obtain
\begin{align}
    \mathcal{S} = \lim_{t_F \to + \infty} t_F^{-1} \int_{t_0}^{t_F} dt\, \left( \mathcal{L}^{TRB}[\varphi] - \mathcal{L}^{TRB}[\varphi]_R \right)
\end{align}
where for computing $\mathcal{L}^{TRB}[\varphi]_R$ we apply the time-reversal transformation to $\mathcal{L}^{TRB}[\varphi]$.
After standard manipulation we obtain
\begin{align}
    \mathcal{S} &= \frac{\tau^2}{2 D} \left \langle \dot{\varphi}_k \dot{\varphi}_q \dot{\varphi}_l G_{kql}\right \rangle \\ \label{eq:sigma2}
G_{kql} &\equiv \frac{\partial^3 H_{LG}}{\partial \varphi_k \partial \varphi_q \partial \varphi_l } = \frac{\partial^3 H_I}{\partial \varphi_k \partial \varphi_q \partial \varphi_l}\; .
\end{align}
Once we perform the inverse Fourier transform, the total entropy production rate $\mathcal{S}$ can be written as an integral over the space of a density of entropy production rate $\sigma$ as follows \cite{PhysRevE.105.044139}
\begin{align} \label{eq:epr_new}
    \mathcal{S} &= \int d\xx \, \sigma(\xx) \\ \nonumber 
    \sigma(\xx) &\equiv \frac{\tau^2}{2 D}\left\langle \dot{\varphi}^3 \frac{\delta^3 H_I}{\delta \varphi^3} \right\rangle \; ,
\end{align}
the integral over the space in (\ref{eq:epr_new}) is bounded below by the ultraviolet cutoff $\Lambda$ so that we avoid any ultraviolet divergence.

\subsection*{Numerical implementation of Model A}
In the case of a non-conserved scalar field $\varphi$, the dynamics can be written as follows
\begin{align} \label{eq:gen_dyn}
    \dot{\varphi} &= F[\varphi] + \psi \\ 
    \dot{\psi} &= \tau^{-1} \psi + \xi 
\end{align}
The noise $\xi(\xx,t)$ satisfies 
$\langle \xi (\xx,t) \rangle =0$ and $\langle \xi(\xx,t) \xi(\xx^\prime,s) \rangle = 2 D \tau^{-2} \delta(\xx - \xx^\prime) \delta(t-s)$, with $D$ measuring the strength of the noise, 
so that the deterministic force is given by,
as usual, 
\begin{align}
    F[\varphi] = -\frac{\delta H_{LG}}{ \delta \varphi} = \left( \mu \nabla^2 - a - \frac{u}{3} \varphi^2 \right) \varphi \; . 
\end{align}

In order to solve numerically the dynamics of the field in two spatial dimensions, we discretize the fields $\varphi \to \varphi_{i,j}$ ($F[\varphi] \to f_{i,j})$, $\psi \to \psi_{i,j}$, and the noise $\xi\to \xi_{i,j} $ 
on a square lattice $L \times L$ of lattice spacing $\Delta x = \Delta y =1$. For evaluating the Laplacian operator, we adopt the standard scheme
\begin{align}
g_{i,j} \equiv    \nabla^2 \varphi_{i,j} &= \frac{1}{\Delta x^2} \left[ \varphi_{i+1,j} + \varphi_{i-1,j} + \varphi_{i,j+1} + \varphi_{i,j-1} - 4\, \varphi_{i,j}\right] \; .
\end{align}
The numerical integration is thus performed using the Euler scheme that brings to (where, for the generic observable $\mathcal{O}$, we adopt the notation $\mathcal{O}_{i,j}^t \equiv \mathcal{O}_{i,j}$)
\begin{align}
    \varphi_{i,j}^{t+1} &= \varphi_{i,j} + (f_{i,j} + \psi_{i,j}) \Delta t \\ 
    \psi_{i,j}^{t+1} &= \psi_{i,j} +  h_{i,j} \Delta t \\
    f_{i,j} &\equiv g_{i,j} - a \varphi_{i,j} - u \varphi_{i,j}^3 \\ 
    h_{i,j} &\equiv -\frac{1}{\tau} \eta_{i,j} + \xi_{i,j} \; ,
\end{align}
where the noise $\xi_{i,j}$ has zero mean and variance $\frac{D}{\tau^2 \Delta t}$. In our numerical simulations, we set $D=u=1$. 

\subsection*{Numerical implementation of Model B}
In the case of Model B, the equations of motion to solve numerically are
\begin{align}
    \dot{\varphi} &= - \boldsymbol{\nabla} \cdot \boldsymbol{J} \\ 
\boldsymbol{J} &\equiv - \boldsymbol{\nabla} \frac{\delta H_{LG}}{\delta \varphi} + \boldsymbol{\psi} \\ 
\dot{\boldsymbol{\psi}} &= -\tau^{-1} \boldsymbol{\psi} + \boldsymbol{\xi} \; ,
\end{align}
where we have introduced  the vectorial Ornstein-Uhlenbeck field $\boldsymbol{\psi}$
that is the source of correlated noise in the model.
In two spatial dimensions, one has $\boldsymbol{J}=(J_x,J_y)$, and $\boldsymbol{\psi}=(\psi_x,\psi_y)$. The numerical solution in two dimensions can be addressed as follows
\begin{align}
    \varphi_{ij}^{t+1} &= \varphi_{ij} - \left[g_{ij}^x + g_{ij}^y \right] \Delta t \\ 
    (\psi_{ij}^{x,y})^{t+1} &= \psi_{ij}^{xy} - \frac{1}{\tau} (\psi_{ij}^{x,y}) \Delta t  + \xi_{ij}^{xy} \Delta t \\
    g_{ij}^{x,y} &\equiv \partial_{x,y} J_{ij}^{x,y} \\
    J_{ij}^{x,y} &\equiv -h_{ij}^{x,y} + \psi_{ij}^{x,y} \\ 
    h_{ij}^{x,y} &\equiv \partial_{x,y} F_{i,j} \\
    F_{i,j}      &\equiv \frac{\partial H_{LG}}{\partial \varphi_{ij}} \; .
\end{align}

\section*{Numerical Simulations} 
\paragraph*{\underline{Parameters.}} 
We perform numerical simulations for $N_t=10^6$ integration steps with a time step $\Delta t=10^{-2}$, sampling stationary configurations every $10^2$ steps. To reach stationary configurations, we consider from $2 \time 10^4$ up to $10^6$ initial steps.  
To compute the phase diagram of Model A, we have performed numerical simulation for linear lattice size $L=80$ (the system counts $N = L \times L$ lattice sites), $\tau=[0.05,0.1,0.2,0.4,0.6,0.8,1,2,4,6,8,10,20,40,80]$, and $a\in[-1.4,0.25]$, sampling $N_a=33$ values separated by $\Delta a=0.05$. The initial configuration of the field $\varphi^0 \equiv \varphi_{i,j}(0)$ has been drawn random from a uniform distribution with $\varphi^0 \in [-1.5,1.5]$. 
In Model A, we also consider averages over $N_s=50$ independent runs.
In the case of Model B, the phase diagram has been established considering an initial drop of radius $R=30$ in the phase $\varphi^0_{+} = 1.5$  with  $\varphi^0 \in [-1.5,1.5]$ randomly and uniformly distributed outside. In the case of Model B, we consider $L=100$, $\tau=[0.01,0.02,0.04,0.06,0.08,0.1,0.2,0.4,0.6,0.8,1,2,4,6,8,10,20,40,60,80]$, and $a\in[-1.4,0.5]$, sampling $N_a=19$ values separated by $\Delta a=0.1$. In Model B, we study the extension of the phase-separated region by considering different initial radii, $R=10,15,20,24,28,30,32,34,36,38,40,42,44,46,48,50,52,54$, and $\tau=0.1,1,10,40$ (the values of $a$ are the same considered for computing $a_c(\tau)$). The error bars in the phase diagram reflect the grid spacing we employed for spanning the two control parameters $a$ and $\tau$.

\begin{figure*}[!t]
\centering\includegraphics[width=1.\textwidth]{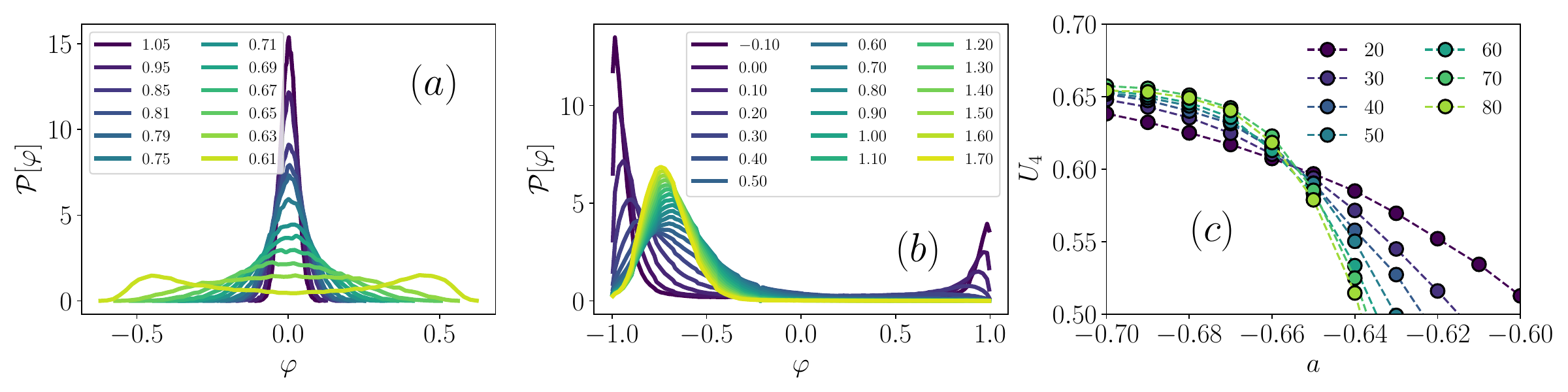}
\caption{
(a) Probability distribution function $\mathcal{P}[\varphi]$ by varying $a$ (see the legend that indicates $a - a_c$) for $\tau=1$ (here $L=80$) in the case of Model A. The distribution is single-peaked (centered around $\varphi=0$) in the disordered phase, it develops a double peak in the ordered phase. (b) $\mathcal{P}[\varphi]$ in the case of Model B where, because of density conservation, $\mathcal{P}[\varphi]$ is centered around $\varphi_a$ in the disordered, homogeneous, phase. (c) Binder parameter $U_4$ for different system sizes (see legend) and $\tau=1$ (Model A). The crossing of $U_4$ indicates a scale-invariant region corresponding to the critical point.}
\label{fig:fig1SI}      
\end{figure*}

\paragraph*{\underline{Phase Diagram and Finite-Size Scaling.}} We explore the phase diagram of the model through the probability distribution function of the field $\mathcal{P}[\varphi]$ that is obtained by computing the following histogram
\begin{align}
    \mathcal{P}[\varphi] &\equiv \left\langle \delta\left[ \varphi(t) - \varphi \right] \right\rangle \\
    \varphi(t) &\equiv \frac{1}{N} \sum_{i,j} \varphi_{i,j}^t
\end{align}
the angular brackets indicate the time average over stationary trajectories.
To identify disordered and ordered phases, we look at the shape of $\mathcal{P}[\varphi]$. Where $\mathcal{P}[\varphi]$ is single-peaked, the system is in a disordered (homogeneous) phase. Ordered (symmetry broken) phases are identified by a double-peak in the $\mathcal{P}[\varphi]$. The typical behavior of $\mathcal{P}[\varphi]$ crossing the transition in the two models is shown in Fig. (\ref{fig:fig1SI}a,b).
To verify whether or not the universality class is compatible with Ising $2d$ for $\tau\neq0$, we performed a Finite Size Scaling analysis \cite{amit2005field} at $\tau=1$.
To do that, we performed numerical simulations at different system sizes $L=40,50,60,70,80$. We compute the Binder parameter defined as \cite{binder1981finite} that is
\begin{align}
    U_4(L,a,\tau) &\equiv 1 - \frac{\langle \varphi^4 \rangle}{3 \langle \varphi^2 \rangle^2} \\ 
    \langle \varphi^m \rangle &\equiv \int d\varphi \mathcal{P}[\varphi] \varphi^m \; .
\end{align}
From the crossing of $U_4$ for different system sizes as a function of $a$, we obtain the critical value $a_c(\tau=1)$ (we report in Fig. (\ref{fig:fig1SI}c) the behavior of $U_4$ for different system sizes). The consistency of the critical exponent $\nu$ has been checked by the data collapse of $U_4$vs$(a-a_c)L^{1/\nu}$. That is justified under the Finite-Size Scaling {\it ansatz}
that states for a generic observable $\mathcal{O}$ measured around the critical region in a system of linear size $L$ 
\begin{align}
    \mathcal{O} = L^{x_\mathcal{O} / \nu} \left[ F_\mathcal{O}(L \xi^{-1}) + O(L^{L^{-\omega}},\xi^{-\omega})\right]
\end{align}
with $\xi$ the correlation length that diverges as $|a-a_c|^{-1/\nu}$ at the critical point ($F_\mathcal{O}$ is the finite-size scaling function, $x_\mathcal{O}$ the critical exponent of the observable, and $\omega$ determines the sub-leading correction to the scaling). Once we (i) ignore sub-leading corrections, and (ii) assume $\xi \to \infty$, i.e., $\xi / L \sim 1$ in the case of a finite size system, we obtain
\begin{align}
    \mathcal{O} = L^{x_\mathcal{O} / \nu} \hat{F}_\mathcal{O}(L^{1/\nu} (a- a_c)) \; , 
\end{align}
and thus we obtain the scaling collapse once we plot  $L^{-x_\mathcal{O}/\nu} \mathcal{O}$ vs $L^{1/\nu} (a-a_c)$.
The anomalous dimension $\eta$ is thus obtained by looking at the scaling of the susceptibility $\chi$ defined as
\begin{align}
    \chi \equiv N \langle \left[ \varphi - \langle \varphi \rangle \right]^2 \rangle \; .
\end{align}
Once we compute the susceptibility for different system sizes $\chi_L$, 
the exponent is obtained by looking at the scaling 
plot $ \chi_L (a_c) \equiv \chi_L^{peak}$ vs $L$, with $\chi_L^{peak}$ the peak value of the susceptibility.

\begin{figure*}[!t]
\centering\includegraphics[width=1.\textwidth]{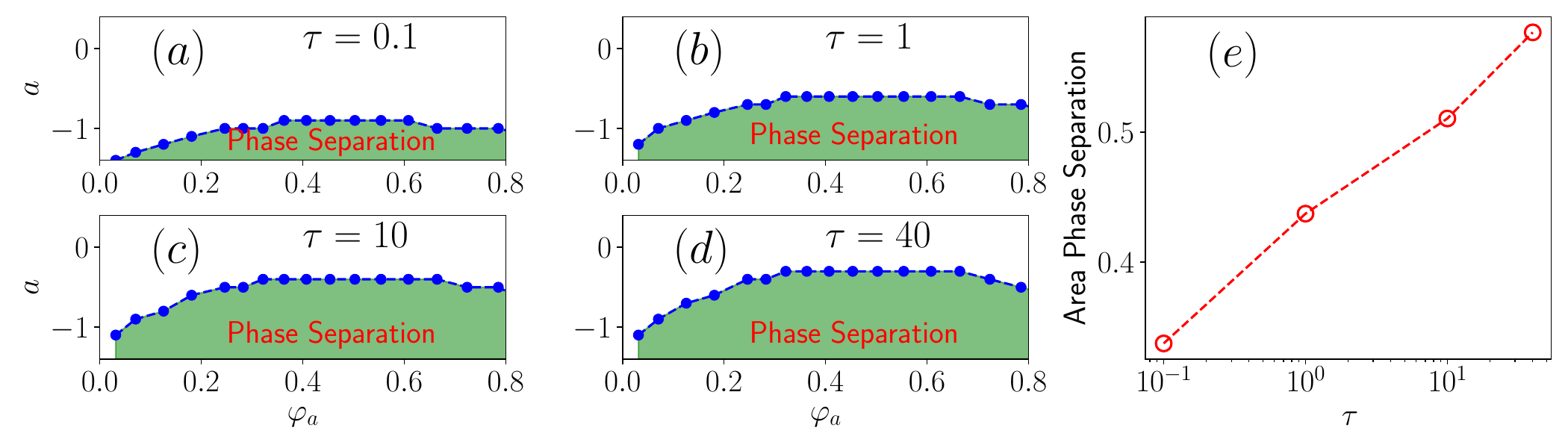}
\caption{
(a)-(d) Phase diagrams of Model B for different values of $\tau$ (see legend).(e) Area of the phase separation region as a function of $\tau$.}
\label{fig:fig2SI}      
\end{figure*}

\paragraph*{\underline{Phase Separation.}}
To study the morphology of phase separation, we performed numerical simulations of Model B starting from two different initial conditions: (i) A drop of radius $R$ of one phase embedded into a random background, and (ii) Random initial configurations with different average density $\varphi_a \equiv \int d\xx \varphi(\xx,0)$ of one of the two phases. To implement the second condition, at the beginning of the simulation, we extract random lattice sites $(i,j)$  and set the field $\varphi_{i,j}=\varphi^0_+$ up to reach the desired $\varphi_a$ value.

\paragraph*{\underline{Microphase Separation.}}
To study microphase separation we have performed numerical simulations for system sizes $N=80^2,120^2,160^2,200^2,240^2$. The other parameters are set at $a=-3$, $\tau=0.5$, and $\rho_a=0.01$. We also studied how $D$ changes microphase by performing numerical simulations for $N=200^2$, $\tau=0.5$, $a=-3$, and $D=0.1,0.2,0.3,0.4,0.5,0.6,0.7,0.8,0.9$. To perform the cluster analysis we adopt the Python package skimage.

\begin{figure*}[!t]
\centering\includegraphics[width=1.\textwidth]{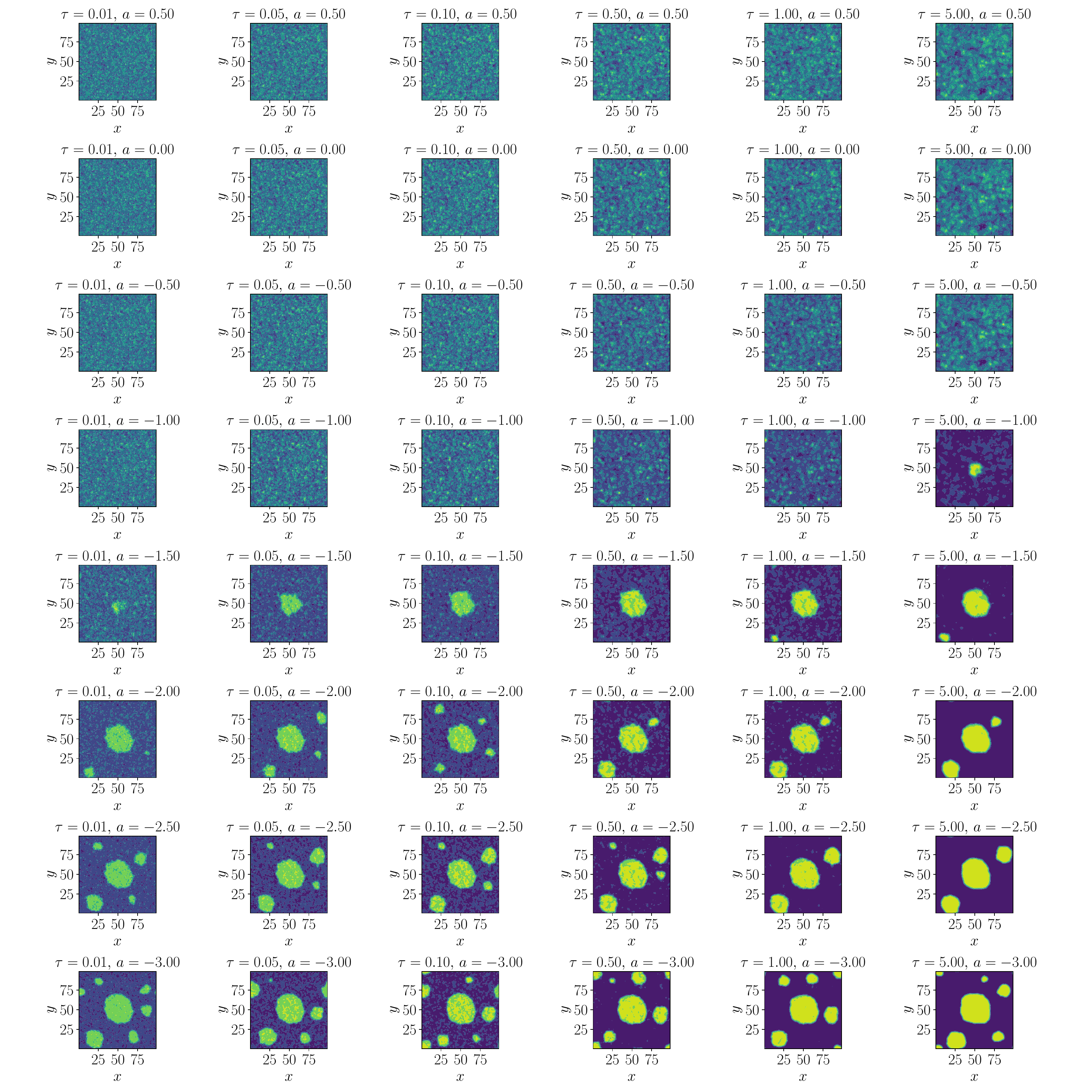}
\caption{
Representative stationary configurations of Model B have been obtained starting from an initial drop of radius $R=10$ (corresponding to $\varphi_a=0.01$). The system size is $L=100$.}
\label{fig:fig3SI}      
\end{figure*}

\begin{figure*}[!t]
\centering\includegraphics[width=1.\textwidth]{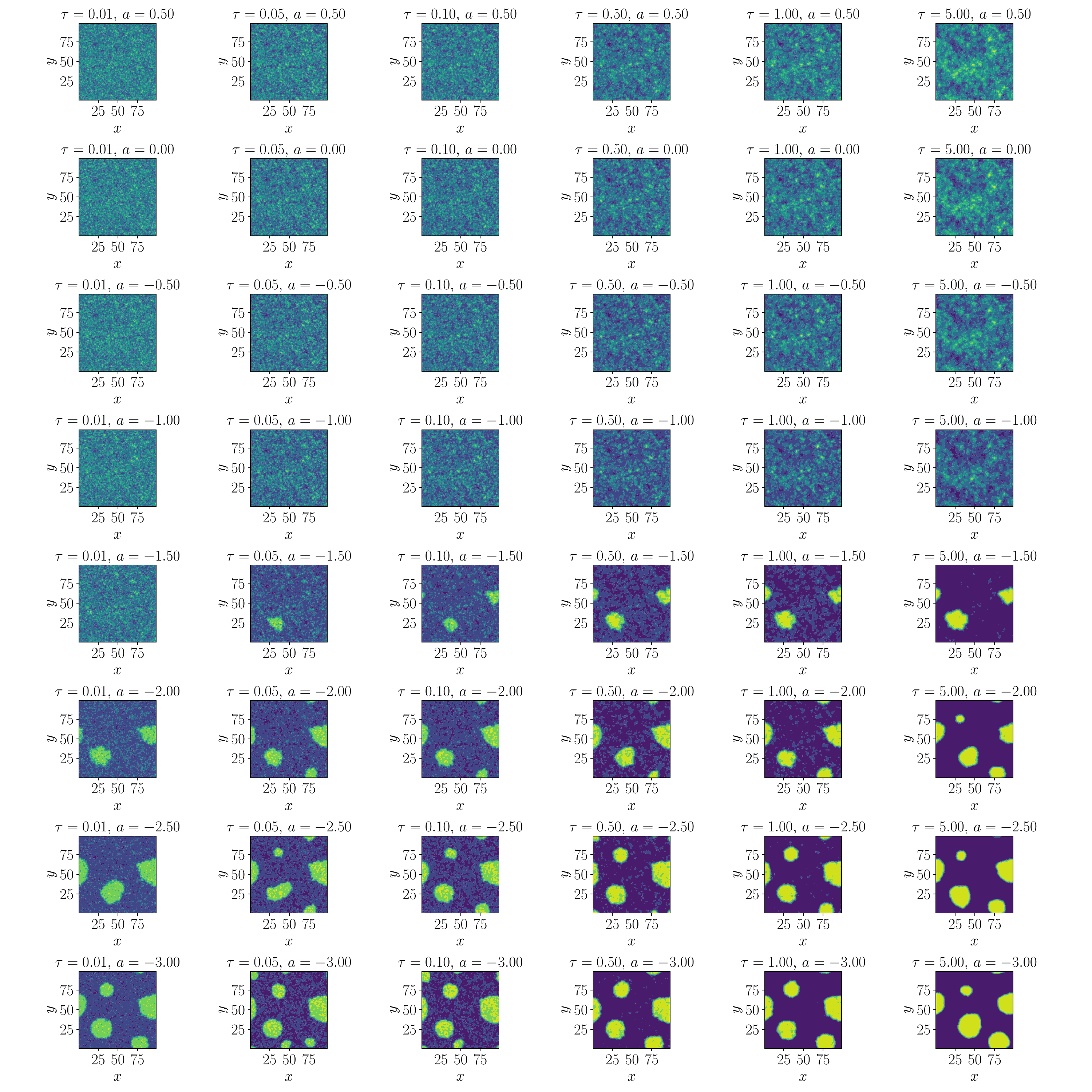}
\caption{
Representative stationary configurations of Model B when the initial configuration is taken as a random configuration at $\varphi_a=0.03$. The system size is $L=100$.}
\label{fig:fig4SI}      
\end{figure*}

In Fig. (\ref{fig:fig3SI}) we report the stationary configurations obtained for different values of $\tau$ and $a$ in the case of an initial drop of radius $R=10$ (with lattice size $L=100$ so that $\varphi_a=0.03$). 
As one can see, the initial drop divides into a microphase of many drops. The same happens in the case of random initial configurations, as shown in Fig.  (\ref{fig:fig4SI}).

\paragraph*{\underline{Entropy Production Rate.}}
We provide a numerical estimate of the entropy production rate by employing (\ref{eq:epr_new}) so that, in a system of linear size $L$ counting $N=L^2$ lattices sites, we can define
\begin{align}
    \mathcal{S}_L &\equiv \frac{\tau^2}{2 D}\sum_{i,j} \langle \sigma_{i,j} \rangle \\ 
    \sigma_{i,j} &\equiv u \dot{\varphi_{i,j}}^3 \varphi_{i,j} \; .
\end{align}
We compute $\mathcal{S}$ in the case of Model B with parameters $\tau=D=1$, and systems size $L=60$ considering as initial condition a drop of radius $R=20$. To perform the average of stationary configurations of the field, we wait $10^6$ time-steps before computing the time average
The stationary time average is computed after $10^6$ initial time steps over a total of $10^7$ steps.
 \section*{Microscopic model with quenched and annealed noise}
We consider a set of Active Ornstein-Uhlebneck particles in $d$ spatial dimensions interacting through a two-body potential whose dynamics is described by a set of Langevin equations that counts two noises: a standard thermal noise $\xi_i\equiv \boldsymbol{\xi}_i$,  representing the interaction with a thermal bath at temperature $T$, and an Active Noise $\psi_i\equiv \boldsymbol{\psi}_i$ of noise-strength $D$ and persistence time $\tau$, representing the self-propulsive mechanisms. We indicate with $x_i \equiv \boldsymbol{x}_i$ the position of particle $i$ in $d$ spatial dimensions, with $i=1,..,N$. $V_{ij} \equiv V(x_i - x_j)$ is the two-body potential. Finally, we indicate with $\nabla_i\equiv \boldsymbol{\nabla}_{\boldsymbol{x}_i}$. The equations of motion in the overdamped regime are (we set the mobility $\mu=1$)
\begin{align}
    \partial_t x_i = - \sum_j \nabla_j V_{ij} + \xi_i + \psi_i \; .
\end{align}
with the statistical features of the two Gaussian noises specified by (the Greek letters indicate the cartesian components)
\begin{align}
    \langle \xi_i^\mu(t) \rangle &= 0, \; \; \, \langle \xi_i^\mu(t) \xi_j^\nu(s) \rangle = 2 T \delta_{ij} \delta(t - s) \delta^{\mu \nu} \; \\  
    \langle \psi_i^\mu(t) \rangle &= 0, \; \; \, \langle \psi_i^\mu(t) \psi_j^\nu(s) \rangle = 2 D \delta_{ij} K(t - s) \delta^{\mu \nu} \; .
\end{align}
in the case of interest, the memory kernel $K(t)$ is 
\begin{align}
    K(t) &= \frac{1}{\tau} \, e^{-|t| / \tau} \; .
\end{align}
To show how the memory kernel $K(t)$ still characterizes the dynamics on the mesoscopic scale, we now perform a coarse-graining following \cite{dean1996langevin}.
We emphasize that the presence of the thermal noise that acts as an annealed variable plays a crucial role in our derivation of the equations for the slow degrees of freedom. In particular, we are going to apply the Ito lemma that can be employed because we are considering active dynamics as an external quenched random field with respect to the fast fluctuation introduced by the thermal noise. 
We start with introducing the density field $\rho(x,t)$ defined as
\begin{align}
    \rho &\equiv \rho(x,t) = \sum_i \rho_i \\ 
    \rho_i &\equiv \rho_i(x,t) = \delta \left[ x_i(t) - x \right]
\end{align}
once we introduce the local density field $\rho_i$, a generic function $f(x_i)$ can be written as follows
\begin{align}
    f(x_i) = \int dx \, \rho_i f(x)
\end{align}
Considering $\psi_i$ as a quenched variable with respect to the instantaneous time scale of $\xi_i$, we can expand the stochastic differential equation using Ito calculus so that the time derivative of $f$ becomes
\begin{align}
    \frac{df}{dt} = \int dx \, \rho_i \left[ \nabla f \cdot (\xi_i + \psi_i) - \nabla f \cdot  \sum_j V_{ij} + T \nabla^2 f \right]
\end{align}
once we integrate by parts we get
\begin{align}
    \frac{df}{dt} &= \int dx \, f(x) \frac{\partial \rho_i}{\partial t} \\ 
    \partial_t \rho_i &\equiv T \nabla^2 \rho_i + \nabla_i \cdot \left( \rho_i \sum_j V_{ij}\right) - \nabla_i \cdot \left(\rho_i \xi_i \right) - \nabla_i \cdot \left(\rho_i \psi_i \right)
\end{align}
once we sum over $i$ and perform standard manipulations (see Ref. \cite{dean1996langevin} for details), we finally obtain 
the equation of motion for the density field 
\begin{align} \label{eq:rho_eq}
    \partial_t \rho &= \nabla \cdot  \left(\rho \frac{\delta G}{\delta \rho}\right) + \nabla \cdot (\rho^{1/2} \eta_T)+ \nabla \cdot (\rho^{1/2} \eta_A) \\ 
    G[\rho] &\equiv \frac{1}{2} \int dxdy\, \rho(x,t) V(x,y) \rho(y,t) + T \int dx \, \rho(x,t) \log \rho(x,t) \; .
\end{align}
We thus arrive at a stochastic equation driven by two noise fields,
with $\eta_T$ a thermal-like noise field
\begin{align}
    \langle \eta_T^\mu(x,t) \rangle = 0 \; , \;\; \langle \eta_T^\mu(x,t) \eta_T^\nu(y,s)\rangle = 2 T \delta^{\mu \nu} \delta(t-s) \delta(x-y) 
\end{align}
while the active noise field $\eta_A$ is still exponentially correlated on the time-scale $\tau$
\begin{align}
    \langle \eta_A^\mu(x,t) \rangle = 0 \; , \;\; \langle \eta_A^\mu(x,t) \eta_A^\nu(y,s)\rangle = 2 D \delta^{\mu \nu} K(t-s) \delta(x-y) \; .
\end{align}
%\bibliography{mpbib}
%\bibliographystyle{rsc}

\end{document}